\documentclass[%
reprint,
a4paper,
amsmath,
amssymb,
aps, 
physrev,
floatfix,
]{revtex4-2}

\usepackage{graphicx} 
\graphicspath{{figures/}} 
\usepackage[dvipsnames,svgnames*,x11names*]{xcolor}

\usepackage{hyperref}
\hypersetup{
colorlinks=true,
linkcolor=Maroon,
filecolor=Maroon,
citecolor=Blue,
urlcolor=Blue}
\usepackage{enumerate}
\usepackage{booktabs}
\usepackage{longtable}

\usepackage[ruled,vlined,linesnumberedhidden]{algorithm2e}
\usepackage{bm}
\usepackage{amsmath,amssymb}

\usepackage[capitalise]{cleveref}

\newcommand{\parencite}[1]{\citep{#1}}

\newcommand{\FP}{P^t_\alpha}

\newcommand{\M}{\mathcal{M}}
\newcommand{\Mk}{\mathcal{C}_k}
\newcommand{\Mi}{\mathcal{C}_i}
\newcommand{\Mj}{\mathcal{C}_j}

\newcommand{\dist}{\pi}
\newcommand{\statdist}{\dist^\infty}

\newcommand{\qel}{Q}
\newcommand{\tel}{T}
\newcommand{\Q}{\mathbf{\qel{}}}
\newcommand{\T}{\mathbf{\tel{}}}
\newcommand{\Qhat}{\widehat{\mathbf{\qel{}}}}
\newcommand{\That}{\widehat{\mathbf{\tel{}}}}

\newcommand{\phioc}{\phi_\alpha}
\newcommand{\phiocstar}{\phi_{\alpha^*}}
\newcommand{\dhat}{\widehat{\mathbf{d}}}
\newcommand{\Rhat}{\widehat{\mathbf{R}}}
\newcommand{\gammahat}{\widehat{\gamma}}

\newcommand{\TV}{{\rm TV}}

\newcommand{\MAE}{{\rm MAE}}

\begin{document}

\title{
Control-oriented cluster-based reduced-order modelling 
}

\author{P. Olivucci}
\author{D. E. Rival}
\author{R. Semaan}
\email{richard.semaan@zeiss.com}
\altaffiliation{currently at Zeiss SMT}
\affiliation{Institut für Strömungsmechanik, TU Braunschweig, 38108 Braunschweig, Germany}

\begin{abstract} 

This work addresses the challenge of learning reduced-order models (ROMs) capable of generalizing to unobserved dynamical regimes across unseen control parameters. 
We introduce the Control-oriented Cluster-based Network Model (CNMc), a framework for synthesizing reduced-order dynamics at held-out operating conditions without requiring simulation data at those conditions. 
While the traditional Cluster Network Model (CNM) is limited to observed regimes, CNMc enables generalization by fitting supervised regression models to the transition probabilities and transition times of the CNM as functions of the control parameter. 
A key enabler is a Procrustes transformation that maps each operating condition's state space to a common coordinate system in which trajectories across all conditions are standardised and shape-aligned, permitting a shared cluster partition to be learned. 
We evaluate CNMc on two canonical fluid dynamics benchmarks — the Lorenz-63 system and a controlled turbulent boundary layer — demonstrating that the predicted statistics at the withheld condition closely match those of a CNM trained directly on test data. CNMc also outperforms the competing interpolation-based CNM approaches under identical conditions. 
These results represent a step toward parameter-aware ROMs suitable for real-time flow control and the acceleration of parametric design studies.
\end{abstract}

\maketitle



\section{Introduction}

Model-order reduction seeks to identify low-dimensional representations that capture the
salient features of high-dimensional dynamical systems. These reduced-order models (ROMs) are
indispensable for gaining physical insight into complex phenomena and act as computationally
efficient surrogates for numerous applications including real-time control, optimization, experiment
planning, data assimilation and accelerating numerical simulations. 
While local ROMs—valid only in the vicinity of a specific equilibrium—are well-established
\cite{Guckenheimer1983}, there is a growing necessity for global ROMs capable of approximating
nonlinear dynamics across the entire state space \cite{Cenedese2022}. 
In fluid mechanics, which serves as the archetypal high-dimensional nonlinear system, the
complexity is compounded by the emergence of distinct dynamical regimes as control parameters (e.g.,
Reynolds number) vary. 
This context motivates the challenge of developing global ROMs capable of handling different
dynamical regimes.  

Traditional data-driven ROMs, such as Dynamic Mode Decomposition (DMD), often rely on linear
evolution operators that are by construction local and struggle to capture globally nonlinear
dynamics.  
While nonlinear alternatives like spectral submanifolds \cite{Cenedese2022} and nonlinear dimension
reduction \cite{DeJesus2023} offer improvements, they typically remain ``anchored'' to a set of
localised dynamical features. 

A parallel class of ``generative ROMs'' shifts the focus from predicting individual trajectories to
reproducing the statistics of the system. These methods are based on approximating the
Frobenius-Perron (FP) operator, the linear operator that advances probability densities rather than
individual states \cite{Froyland2013, Klus2018, Souza2023}. 
By discretizing the state space into a network of probabilistic transitions—a framework known as
Ulam's method \cite{Bollt2013}—one can construct coarse-grained, intrinsically global ROMs. 
Cluster-based Reduced-Order Models (CROMs) exemplify this approach by using clustering algorithms to
learn an efficient state-space discretization directly from data \cite{Burkardt2006}. 
Successes in fluid dynamics, including the Cluster-based Markov Model (CMM) \cite{Kaiser2014} and
the Cluster-based Network Model (CNM) \cite{Fernex2021}, have demonstrated the utility of these
frameworks for flow modeling, control design and data assimilation \cite{Nair2019,Kaiser2024}.  

Despite these advances, a fundamental bottleneck remains: out-of-distribution (OOD) generalization.
Standard CROMs are inherently descriptive; they reproduce dynamics within an observed regime but
lack the mechanism to infer dynamics at unobserved parameter values.
Previous attempts to address OOD modelling have relied on post-hoc weighted interpolation or
``shadowing'' of trajectories generated by single-regime models \cite{Fernex2021a, Iacobello2022}. 
However, these approaches are limited by the requirement of strong structural similarity between the
observed and target regimes and often fail when the flow physics undergoes a significant qualitative
shift. 

In this work, we introduce the Control-oriented Cluster-based Network Model (CNMc) to move beyond
simple interpolation. Rather than shadowing known trajectories, CNMc directly models the evolution
of the statistical transition parameters themselves via self-supervised learning. This allows for
the zero-shot synthesis of reduced-order dynamics at untried control parameters. By treating the
transition network as a parameter-aware operator, CNMc provides a more robust and flexible framework
for OOD generation. We demonstrate the efficacy of this approach on two canonical fluid systems,
highlighting its potential for model-predictive control and accelerated numerical exploration of
parametric nonlinear systems. 

The remainder of this article is organized as follows: \cref{sec:problem_description} defines the
problem statement and the modeling framework; \cref{sec:cnmc} details the CNMc algorithm; and
\cref{sec:numerical_experiments} presents results from numerical benchmarks demonstrating improved
generalization over state-of-the-art baselines.

\section{Methodology}\label{sec:problem_description} 

\subsection{Problem formulation}\label{sec:dynamical_systems}
We consider $N$-dimensional autonomous dynamical systems in which the time evolution of the state
$\mathbf{u}\in\mathbb{R}^N$ is governed by
\begin{equation}\label{eq:dyn_sys}
\frac{\rm d \mathbf{u}}{\rm d t} =  \mathbf{F}_\alpha(\mathbf{u}),
\end{equation}
parametrised by $\alpha$.
For brevity, we take $\alpha\in\mathbb{R}$ throughout this section; the extension to
multi-parameter systems is straightforward.
Equations of the form \cref{eq:dyn_sys} describe natural unforced systems, steadily and
periodically forced systems, and feedback systems \parencite{Guckenheimer1983}.
In the feedback case, the right-hand side of \cref{eq:dyn_sys} can be written explicitly as
\begin{equation}
\mathbf{F}_\alpha(\mathbf{u}) = \mathbf{F}(\mathbf{u}, \mathbf{b}_\alpha(\mathbf{u})),
\end{equation}
where $\mathbf{b}_\alpha(\mathbf{u})$ is a parametric feedback-control law.
To unify its diverse physical meanings, we refer to the parameter $\alpha$ interchangeably
as the ``operating condition'' (OC) or ``regime''.
Representative examples include the Reynolds number in naturally evolving flows, the frequency
of an open-loop periodic forcing, and the gain of a feedback-control policy.

We are primarily concerned with the statistical properties of solutions of \cref{eq:dyn_sys};
specifically, with the evolution of an ensemble of states described by the probability
density function $\dist(\mathbf{u}, t)$ under the action of the vector field $\mathbf{F}_\alpha$.
The inter-temporal conservation of probability is expressed by the Liouville equation
\begin{equation} \label{eq:liouville}
\frac{\partial}{\partial t} \dist(\mathbf{u}, t) =  
-\nabla_\mathbf{u} \cdot \bigl(\dist(\mathbf{u}, t)\, \mathbf{F}_\alpha(\mathbf{u})\bigr),
\end{equation}
which admits the operatorial form
\begin{equation}\label{eq:fp_operator}
\dist(\mathbf{u}, t) = \FP\, \dist^0(\mathbf{u}),
\end{equation}
where $\dist^0$ is the distribution of initial conditions and $\FP$ denotes the Frobenius--Perron
operators, which propagate distributions forward in time \parencite{Lasota1994}.

We restrict attention to ergodic dynamics on a subdomain $\M{}$ of the state space, typically a
lower-dimensional manifold, as is the case for the global attractor of a dissipative dynamical
system such as a turbulent fluid flow \parencite{Holmes2012}. 
Under the ergodic assumption, the long-time statistics become insensitive to initial conditions.
Equations~\cref{eq:liouville,eq:fp_operator} then admit a unique steady-state solution
$\statdist_\alpha(\mathbf{u})$ supported on $\M{}$, the stationary distribution, to which every
non-singular $\dist^0(\mathbf{u})$ converges.

The central modelling objective is to leverage trajectory observations collected at several
operating conditions to predict the reduced-order dynamics at unobserved operating conditions.
The problem is formalised as follows.

Let $\alpha^*$ denote an unobserved operating condition of system \cref{eq:dyn_sys}, and let
$\alpha_1,\ldots,\alpha_M$ be a set of $M$ operating conditions for which observations
$\mathbf{U}_1,\ldots,\mathbf{U}_M$ are available.
Each observation $\mathbf{U}_m = (\mathbf{u}^1,\ldots,\mathbf{u}^T)^\top \in \mathbb{R}^{T \times N}$
for $m=1,\ldots,M$ is a trajectory of the system recorded at $T$ discrete time instants.
The objective is to generate statistically correct trajectories $\mathbf{U}^*$ at the
unobserved condition $\alpha^*$, which amounts to sampling from the conditional model
\begin{equation}\label{eq:generative_model}
\mathbf{U}^* \sim p(\mathbf{U} \mid \alpha=\alpha^*).
\end{equation}
Learning $p(\mathbf{U} \mid \alpha)$ from the observations decomposes into two sub-problems.

The first sub-problem is generating trajectories $\mathbf{U}$ at a fixed value $\alpha=\alpha_m$.
Invoking ergodicity, \cref{eq:generative_model} can be written as the stationary Markov model
\begin{equation}\label{eq:markov_model}
p(\mathbf{U} \mid \alpha_m) = \prod_{t=1}^T p(\mathbf{u}^t \mid \mathbf{u}^{t-1}, \alpha_m),
\end{equation}
which enables simulation of sequences of arbitrary length $T$ once the transition kernel
$p(\mathbf{u}^t \mid \mathbf{u}^{t-1}, \alpha_m)$ has been estimated from $\mathbf{U}_m$.
A CROM approximates the full state $\mathbf{u}$ through a finite number of representative states,
yielding a discrete Markov kernel in the form of a transition probability matrix.
The specific CROM formulation adopted here and its associated kernel are reviewed in detail in
\cref{sec:crom_models}.

The second sub-problem is to endow the trajectory model with the ability to learn the conditional
dependence on $\alpha$. 
This is the subject of the present contribution and is discussed in \cref{sec:cnmc}. 

\subsection{Cluster-based network modeling}\label{sec:crom_models}
Before presenting the proposed method, we briefly review the Cluster-based Network Model (CNM),
which this study buids on. 
CNM is a type of CROM, an approach that captures the dynamics of \cref{eq:dyn_sys} at some $\alpha$
as a random walk on a discrete partition of the state space by learning a discrete kernel for
\cref{eq:markov_model}. 
The partition consists of $K$ Voronoi cells whose centroids serve as representative states for
all points within each cell.
A key feature of CROMs is that this partition is learned directly from data via a clustering
algorithm, which mitigates the curse of dimensionality that affects regular partitions of
high-dimensional spaces \parencite{Bollt2013, Kaiser2017}.
The CNM is a CROM capable of generating trajectories at arbitrary time resolution \cite{Fernex2021}. 

Throughout this section, all subscripts $\alpha$ are dropped as symbols refer to a single operating
condition. 
The CNM models the dynamics as a jump process between cells with deterministic transition times,
and is defined by the following components:
\begin{enumerate}[(a)]
\item A state-space partition of $K$ non-overlapping cells $\Mk$ satisfying $\bigcup_k^K \Mk = \M{}$, 
each with centroid $\mathbf{c}_k$.
The cells $\Mk$ are obtained by applying the $k$-means clustering algorithm to the observed
trajectories $\mathbf{U}$.
\item A direct transition probability matrix $\Q{}\in[0,1]^{K \times K}$ and a transition-time
matrix $\T{}\in\mathbb{R}_{+}^{K \times K}$, whose elements are
\begin{align}\label{eq:Q_and_T}
\qel{}_{ij} &= P(\mathbf{u}^+\in\Mi \mid \mathbf{u}^-\in\Mj), \quad i \neq j,\\
\tel{}_{ij} &= \tfrac{1}{2}\bigl(\mathbb{E}[T_i] + \mathbb{E}[T_j]\bigr),
\end{align}
where superscripts $^-$ and $^+$ denote the current and next cell visited by a trajectory,
and $T_i$ is the holding time in cell $i$.
The entries of $\Q{}$ are the probabilities of transitioning from the current cell to each other
cell, while those of $\T{}$ are the durations of the corresponding transitions.
Both $\qel{}_{ij}$ and $\tel{}_{ij}$ are estimated from trajectory data by counting pairwise
transition events and averaging cell holding times, respectively.
The model can be extended to a higher-order Markov process with a number of delays $L>1$ by
constructing $\Q{}\in[0,1]^{K^{(L+1)}}$ and $\T{}\in\mathbb{R}_{+}^{K^{(L+1)}}$.
This extension can improve predictive accuracy, though at a worst-case memory cost that grows
exponentially with $L$.
\item Interpolators $\psi_{ij}(t)$ between the $i$-th and $j$-th cell centroids, implemented
as piecewise low-order polynomials parametrised by the time $t$.
\end{enumerate}
Trajectories are generated by combining these elements into a sampling step and an interpolation step:
\begin{align}
k^+ &\sim \qel{}_{k k^-}, \label{eq:cnm_step1} \\
\mathbf{u}^{t} &= \psi_{k^-k^+}\!\left( \frac{(t-t_{k^-})\,\Delta t}{\tel{}_{k k^-}} \right). \label{eq:cnm_step2}
\end{align}
The first equation advances the discrete process from the current cell $k^-$ to the destination
cell $k^+$; the second continuously interpolates between their respective centroids over the
transition time at temporal resolution $\Delta t$.
Together, \cref{eq:cnm_step1,eq:cnm_step2} define the CNM kernel within the general framework of
\cref{eq:markov_model}.
The reader can refer to the literature for an exhaustive account of the CNM algorithm
\cite{Fernex2021}.

\subsection{Control-oriented CNM}\label{sec:cnmc}
The proposed control-oriented Cluster-based Network Model (CNMc) extends CNM to unseen operating
conditions by learning the dependence of the state-space partition and transition properties on
$\alpha$ through supervised regression models.
These models predict the discrete states and transitions at an unseen OC, which are then used to
generate trajectories in the same fashion as standard CNM.

A central challenge in this approach is that, to learn the dependence of the transition properties
on $\alpha$, the discrete cells must remain consistent across all OCs.
In general, however, the dynamics at each $\alpha$ evolves on a distinct manifold $\M{}_\alpha$,
which also constitutes the support of the stationary distribution $\statdist_\alpha$.
A naive partitioning strategy that clusters all trajectory observations simultaneously would, in
most cases, be suboptimal: cell placement would either fail to resolve all OCs with adequate
fidelity or would require an impractically large $K$.
Partitioning each OC independently would be more efficient but would necessitate aligning, or
matching, cells across partitions in order to interpolate the transition properties.
The compromise adopted here is to map all OCs to a set of common latent coordinates $\mathbf{u}'$ in
which the observations are approximately aligned, and to perform partitioning in $\mathbf{u}'$.

More precisely, we consider the family of transformations $\phioc(\mathbf{u})$ that map the
stationary distribution at any $\alpha$ onto that of a reference operating condition $\alpha_0$.  
If $\statdist_0(\mathbf{u})$ is the stationary distribution at $\alpha_0$, then the change of
variables $\mathbf{u}'= \phioc(\mathbf{u})$ is chosen such that $\statdist_\alpha(\mathbf{u}') =
\statdist_0(\mathbf{u})$. 
If $\phioc$ were known, $\mathbf{u}'$ would provide a universal state space for all OCs.
In general, $\phioc$ is a nonlinear map, and learning it from limited data can be severely
ill-posed. 

To address this, we write $\phioc$ as the composition of an affine and a nonlinear transformation,
\begin{equation}\label{eq:map_expansion}
\phioc(\mathbf{u}) = \widetilde{\phi}_\alpha\!\left( \mathbf{B}_\alpha \mathbf{u} + \mathbf{d}_\alpha \right),
\end{equation}
where $\mathbf{B}_\alpha$ is an orthogonal matrix and $\widetilde{\phi}_\alpha$ is the
non-orthogonal, nonlinear residual. 
The rationale for this decomposition is that in the linearly transformed coordinates
$\mathbf{u}' = \mathbf{B}_\alpha \mathbf{u} + \mathbf{d}_\alpha$, the orthogonal component of the
mapping is resolved analytically, leaving the clustering procedure to account only for the
residual variance arising from non-orthogonal and nonlinear ``shape changes''.
Only the orthogonal component is learned from data.
A further practical advantage is that the inverse of an orthogonal transformation is trivially
computed as its transpose.
The specific form of \cref{eq:map_expansion} introduces an inductive bias that may appear
restrictive. 
It can nonetheless be justified as a consistent approximation in the limit of vanishing distance
between training OCs.
When additional system-specific knowledge is available, bespoke transformation classes may be
substituted.

\begin{figure*}[ht!]
  \centering
  \includegraphics[width=\linewidth]{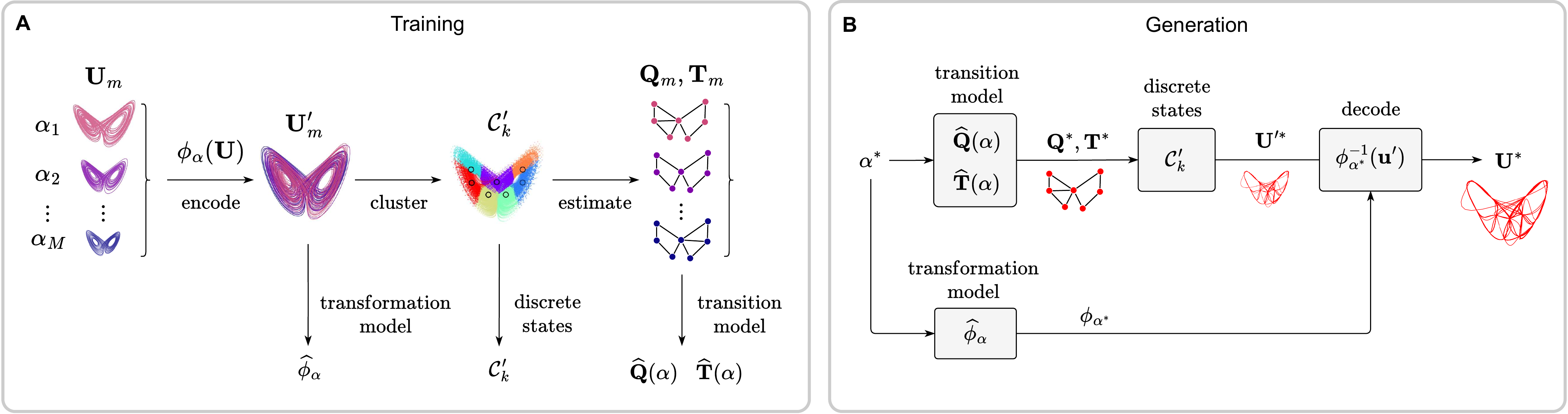}
  \caption{
  {\bf Schematic of the proposed CNMc algorithm.}
  (a) Training stage.
  (b) Inference stage.
  The cluster-based models are represented as graphs, where vertices correspond to cell centroids
  and edges to pairwise transitions.
  }
  \label{fig:cnmc_algorithm}
\end{figure*}

Concretely, the CNMc algorithm comprises four main elements:
\begin{enumerate}[(a)]
\item A family of affine transformations $\phi_\alpha(\mathbf{u})$ that map the state
space to the transformed coordinates $\mathbf{u}'$ in which the observations at every $\alpha_m$ are
standardised and are shape-aligned. 
Specifically, $\phi_\alpha$ is a Procrustes transformation
\begin{equation}\label{eq:procrustes}
\phi_\alpha(\mathbf{u}) = \mathbf{R}_\alpha\,(\gamma_\alpha \mathbf{I})\,\mathbf{u} + \mathbf{d}_\alpha,
\end{equation}
which consists of the composition of a rotation, a uniform scaling, and a translation.
\item A partition $\Mk'$ of $\mathbf{u}'$ with associated cell centroids $\mathbf{c}'_k$.
\item Transition properties $\Q{}_m$ and $\T{}_m$ defined on the partition $\Mk'$.
\item Regression models $\Qhat{}(\alpha)$, $\That{}(\alpha)$ for the transition properties and
$\dhat{}(\alpha)$, $\Rhat{}(\alpha)$, and $\gammahat{}(\alpha)$ for the transformation parameters.
\end{enumerate}

These elements are combined for trajectory generation as sketched in
\cref{fig:cnmc_algorithm}b, and are learnt from trajectory data as sketched in
\cref{fig:cnmc_algorithm}a.
The figure uses the Lorenz-63 system as an illustrative example, which is discussed in detail in
\cref{sec:lorenz_system}. 

The translation vector $\mathbf{d}_m$, the scaling factor $\gamma_m$, and the rotation matrix
$\mathbf{R}_m$ for each $m$-th OC are estimated from the data as follows. 
$\mathbf{d}_m$ is chosen so as to centre the trajectory data at the origin and is thus equal to
minus the trajectory mean state, whose components are $(\mathbf{d}_m)_i = -
\overline{(\mathbf{U}_m)_i}$.  
$\gamma_m$ is equal to the reciprocal of the trajectory standard deviation $\gamma_m =
\left(\overline{\mathbf{U}_m^2}\right)^{-1/2}$ and thus serves to standardise the data.  
Finally, $\mathbf{R}_m$ is chosen as the rotation that optimally aligns (in the least squares
sense) the principal directions of $\mathbf{U}_m$ to those of the first OC $\mathbf{U}_1$, which is
taken as the reference OC (denoted above by $\alpha_0$).
Kabsch's algorithm is used for this purpose \cite{Kabsch1976}. 

The outcome of this procedure is illustrated in \cref{fig:cnmc_algorithm}a, where all transformed
trajectories $\mathbf{U}'_m$ are seen to approximately overlap in the common coordinates
$\mathbf{u}'$.  
The estimated transformation parameters for each OC then become the training data used to fit models
$\dhat{}(\alpha)$, $\Rhat{}(\alpha)$, $\gammahat{}(\alpha)$, collectively denoted by
$\widehat{\phi}_\alpha$.   

The $k$-means clustering step is then performed in $\mathbf{u}'$ on all transformed trajectory data,
yielding $K$ clusters that define a single partition of $K$ cells $\Mk'$ shared across all OCs.

The CNM transition properties $\Q{}_m$ and $\T{}_m$ are subsequently estimated on this common
partition, but independently for each OC.
This is equivalent to learning $M$ independent CNM models, one per OC, on a common partition
optimised to be representative of the full ensemble of operating conditions.

Finally, regression models $\Qhat{}(\alpha)$ and $\That{}(\alpha)$ are fitted to the estimated
transition matrices $\Q{}_m$ and $\T{}_m$.
All entries of $\Q{}$ that are non-zero in at least one training OC are modelled.
The raw output of $\Qhat{}(\alpha)$ is normalised to ensure a valid Markov matrix at all OCs,
i.e.\ so that its entries lie in $[0,1]$ and sum to unity over the first index.

Trajectory generation at an unseen condition $\alpha^*$ proceeds in reverse order.
The transition properties are first predicted from their respective regression models as $\Q{}^* =
\Qhat{}(\alpha^*)$ and $\T{}^* = \That{}(\alpha^*)$; 
a trajectory $\mathbf{U}^{*'}$ is then generated in $\mathbf{u}'$ via the standard CNM procedure; 
and finally the trajectory is mapped back to the original state space by applying the inverse
Procrustes transformation. 





\section{Numerical experiments}\label{sec:numerical_experiments}
The proposed method is demonstrated on two systems: a low-dimensional model problem in
\cref{sec:lorenz_system} and a boundary-layer flow in \cref{sec:controlled_tbl}.
In both cases, the observations consist of numerically integrated trajectories at several
operating conditions, and the experiments are structured as follows.

One OC is withheld as the test condition; the remainder are used to train the CNMc model.
The trained CNMc is then used to generate trajectories at the test OC.
The quality of the CNMc-generated dynamics is primarily assessed by comparing its statistics
to those of a CNM model trained directly on the test OC data, referred to as the reference CNM.
This provides a measure of the out-of-sample generalisation performance of CNMc relative to
its direct in-sample counterpart.
Statistics for both models are computed on the same state-space partition --- cells and
centroids --- namely the reference CNM partition obtained by clustering the test OC data.
Additionally, the statistics of the numerical trajectories at the test OC are also computed on the
same partition and included in the comparison. 
The quantitative criteria used to assess the statistics are the total variation distance (TV)
between discrete stationary distributions and the mean absolute error (MAE) between
autocorrelation functions and power spectra.
The mathematical definitions of these quantities can bye found in \cref{app:statistical_criteria}.

\subsection{Lorenz-63 system}\label{sec:lorenz_system}
The classic Lorenz-63 system is a three-dimensional dynamical system that is a simplified form of
the buoyancy-driven atmospheric convection equations \parencite{Lorenz1963}:
\begin{equation}\label{eq:lorenz}
  \frac{\rm d \mathbf{u}}{\rm d t} = 
  \left( 
    \begin{matrix}
      Pr (u_2 - u_1) \\
      u_1(Ra - u_3) - u_2 \\
      u_1 u_2 - 8 u_3 / 3
    \end{matrix}
    \right).
\end{equation}
When the Prandtl number $Pr$ and the Rayleigh number $Ra$ lie in a certain range, the Lorenz-63
system exhibits chaotic behaviour. 
We fix $Pr=10$ and we study the one-parameter Lorenz dynamics for $Ra$ in $\{30,40,50,60,70\}$,
which define the five operating conditions considered for the system. 
The condition $Ra^*=50$ is reserved as the test case; the remaining four are used for training.

\begin{figure}[h!]
  \centering
  \includegraphics[width=\linewidth]{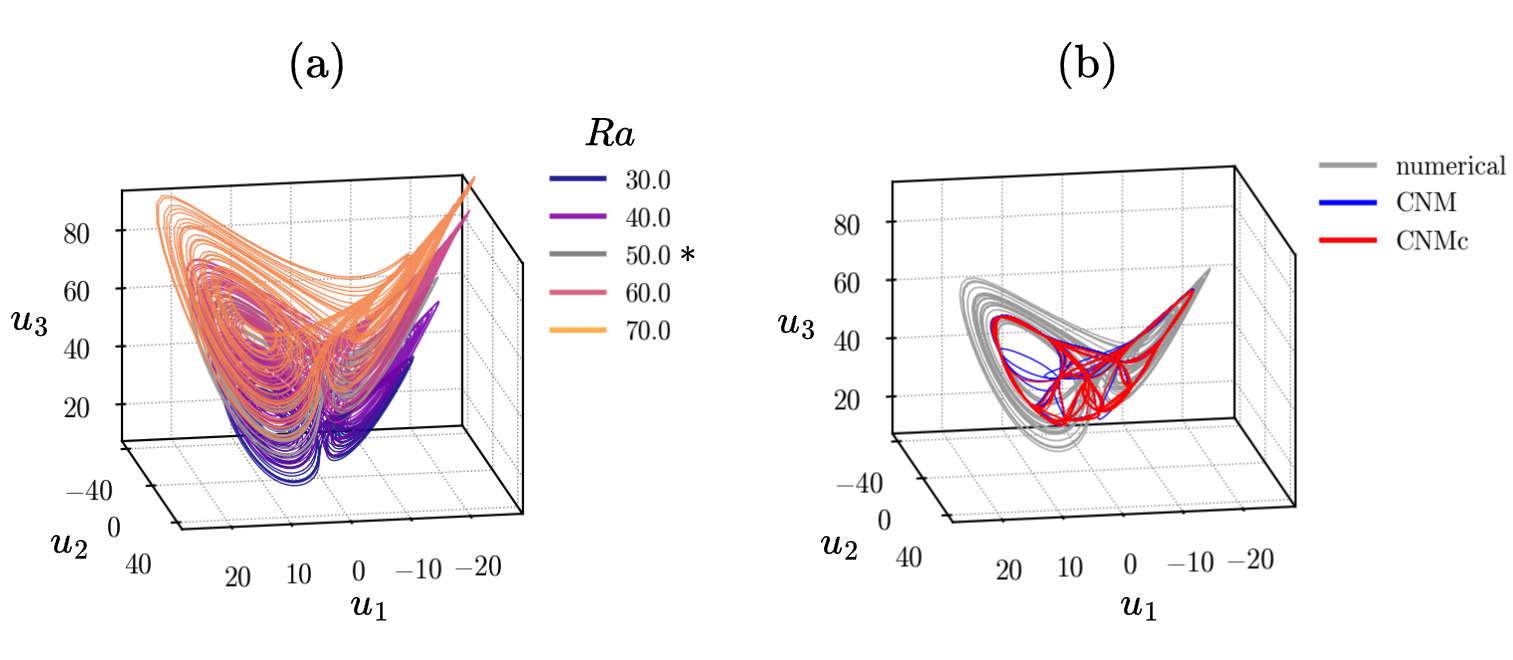}
  \caption{
  {\bf State-space trajectories of the Lorenz-63 system.}
  (a) Numerically integrated trajectories at all values of parameter $Ra$; the test condition
  is indicated by ($^\ast$).
  (b) Generated trajectories at the test condition $Ra^\ast=50$.
  }
  \label{fig:lorenz_data}
\end{figure}

\cref{fig:lorenz_data}a shows the numerically integrated trajectories used as the training and
test data for CNMc.
At all values of $Ra$ considered, the dynamics evolves on a nearly two-dimensional invariant
set, switching chaotically between two lobes.
As $Ra$ increases, the invariant set expands isotropically and the variance along all three
coordinates grows linearly, increasing 2.5-fold from $Ra=30$ to $Ra=70$.
The spectral peak frequency, associated with the periodic dynamics on the lobes, also nearly
doubles over the same range.
Each numerical trajectory comprises $T=500{,}000$ time steps, spanning an integration time of
$8000$ periods of the slowest OC.

CNMc is applied to the Lorenz data with the following hyperparameter settings.
Piecewise linear regression models are chosen for the parameters of $\widehat{\phi}$, and
third-order polynomials with $\ell_1$ regularisation are used for the transition properties. 
These models were selected manually without a systematic model selection study; they are sufficient
to demonstrate the concept, and more sophisticated alternatives are considered in the discussion.  

A number of cells $K=14$ and delays $L=10$ are identified by grid search as the best
combination.
We find that CNMc's accuracy is highly sensitive to $L$ and, to a lesser extent, to $K$.
This is likely attributable to the difficulty of estimating transition probabilities for rare
past histories from finite-length observations.
As the number of possible histories grows exponentially with $L$, a commensurate increase
in data is required to estimate transition probabilities at comparable confidence levels across
all OCs, and to avoid fitting regression models to highly noisy targets.
Further details on the dependence of the CNM--CNMc discrepancy on the hyperparameters are
provided in \cref{app:lorenz_hyperparameter} and appear consistent with data scarcity.

\begin{figure}[h!]
\centering
\includegraphics[width=\linewidth]{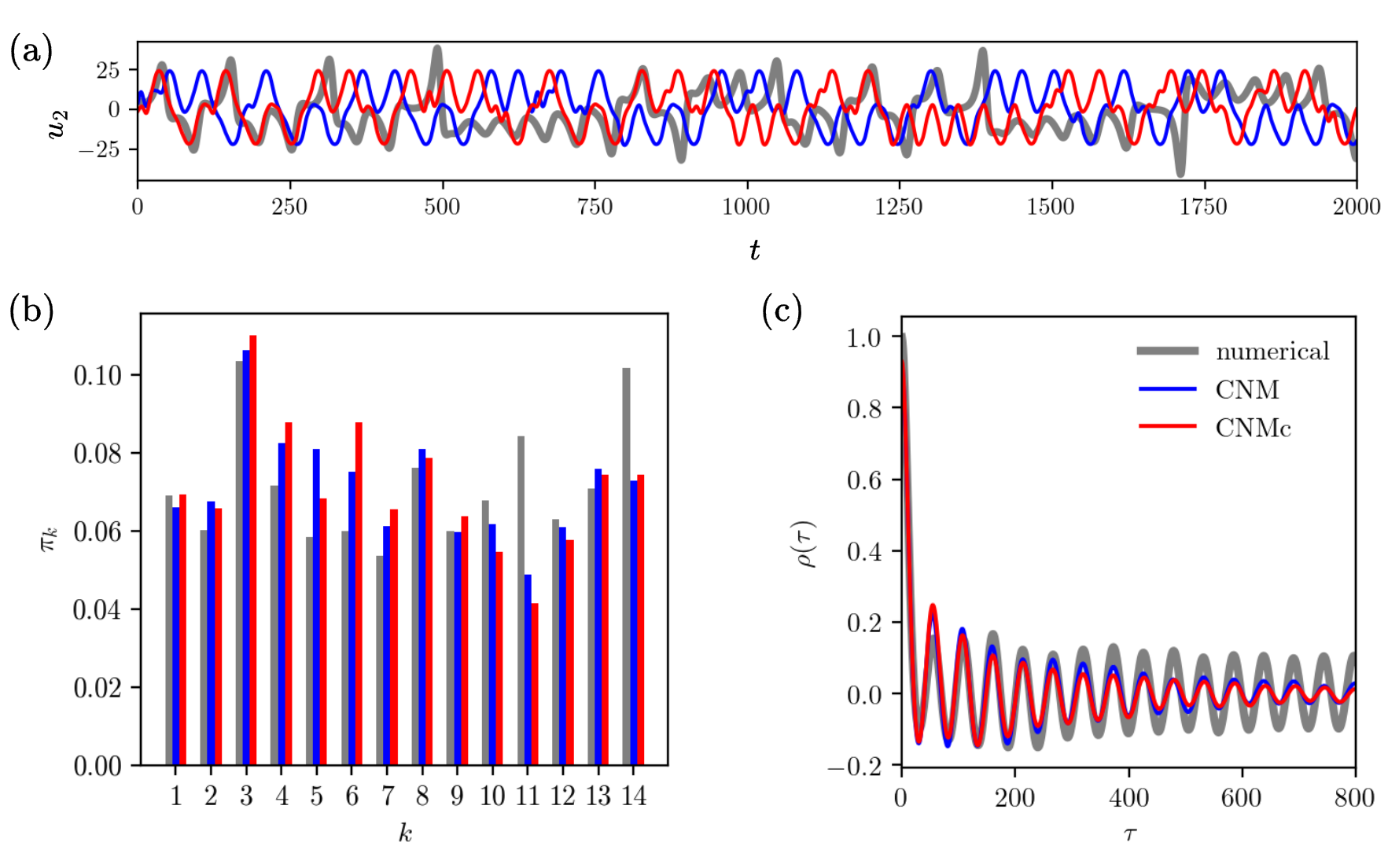}
\caption{
{\bf Lorenz-63 dynamics at the unseen condition $Ra^*=50$.}
CNMc trajectories are compared to the reference CNM and the numerically integrated system:
(a) time-series of the $u_2$ component;
(b) stationary distributions;
(c) auto-correlation functions.
}
\label{fig:lorenz_results}
\end{figure}

The generated dynamics at the unseen condition $Ra^*=50$ are examined in \cref{fig:lorenz_results}.
CNMc is seen to closely recover both the stationary distribution and the temporal
auto-correlation coefficient of the reference CNM fitted to the test condition.
The numerical values of the discrepancy metrics are reported in \cref{tab:errors_models}. 
The departure of both CNM and CNMc from the the true numerical solution reflects the choice of $K$
and $L$.

\begin{figure}[h!]
\centering
\includegraphics[width=\linewidth]{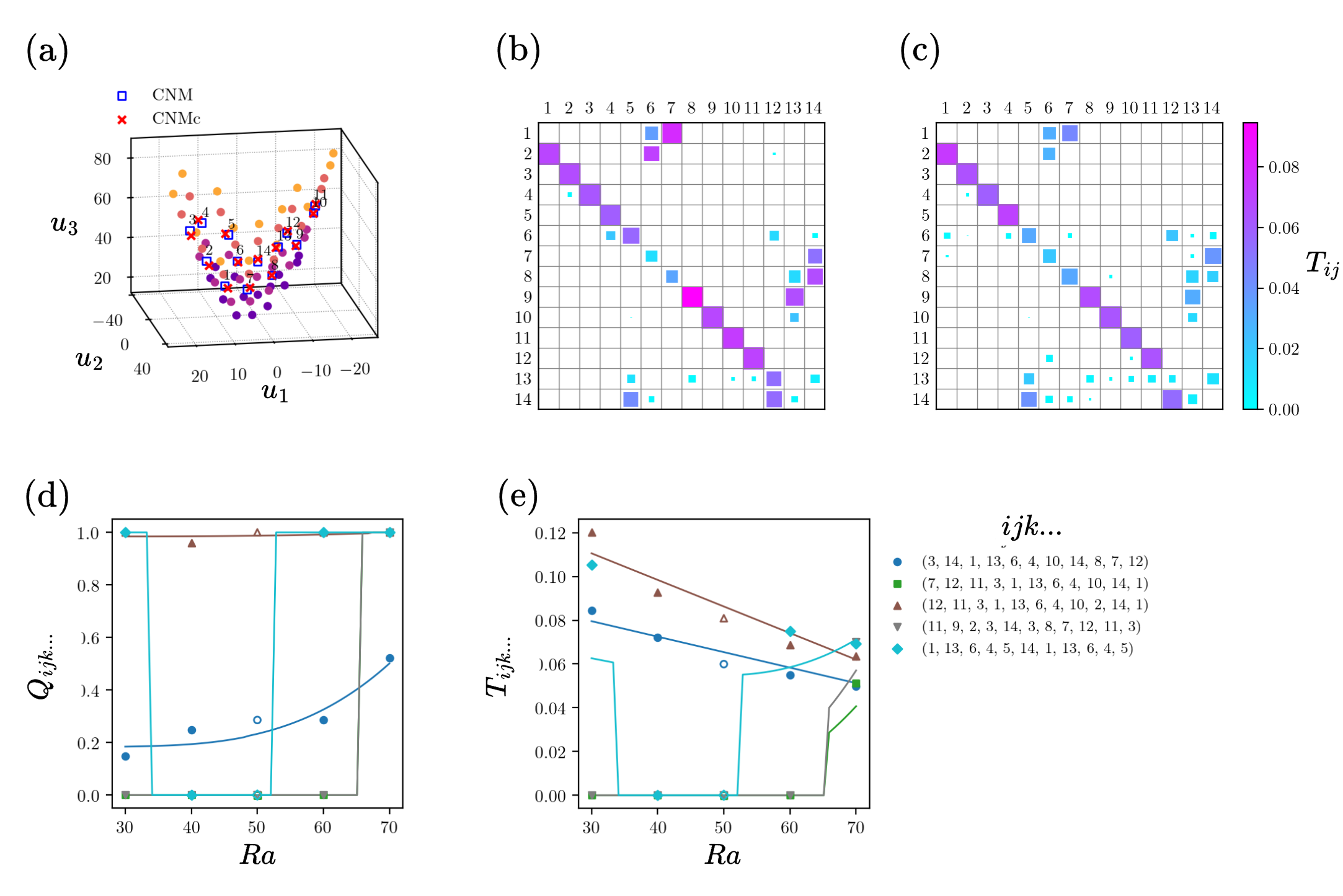}
\caption{
{\bf Elements of the CNMc model, Lorenz-63 data.}
(a) Predicted locations of the discrete states.
(b--c) Hinton plots of the marginalised transition properties for (b) the CNM reference and
(c) the CNMc prediction.
The size of each square represents the transition probability (empty: 0; filled: 1) and the
colour gradient encodes the transition time.
(d) Regression model $\Qhat(b)$ for the transition probabilities (solid lines).
Filled symbols are training data; hollow symbols are the CNM reference at the test OC.
Only five elements are shown; their indices are listed in the legend.
(e) As in (d), but for $\That(b)$.
}
\label{fig:lorenz_cnmc}
\end{figure}

\cref{fig:lorenz_cnmc} compares the learned elements of CNMc to those of the reference CNM 
to illustrate the predictive accuracy of both the cell centroids and the transition properties.
To permit a two-dimensional matrix visualisation of the $L$-dimensional tensor $\Q{}$, the
transition probabilities are marginalised over the trailing $L-1$ indices of $\Q{}$; likewise, the
transition-time matrix $\T{}$ is averaged over the same trailing indices, weighted by the associated
transition probabilities. 
In marginalised form, the reference CNM kernel is well matched by CNMc.

Although a small number of transitions are mispredicted --- either present in CNMc but absent
in the reference, or vice versa --- the majority of the probabilities are correctly placed and
of comparable magnitude, indicating that the overall transition topology has been well captured.
A similar level of agreement is observed for the transition times.

The regression models $\Qhat(b)$ and $\That(b)$ are shown for a representative subset of five
entries in \cref{fig:lorenz_cnmc}d--e.
The noticeable discontinuities in the fifth entry are a consequence of the model suppressing some
transition probabilities around $Ra=40$ to ensure $\Q$ is a stochastic matrix at all OCs. 

To evaluate CNMc's OOD performance more generally, training and testing are also repeated in a
leave-one-out (LOO) fashion for each of the five OCs. 
The hyperparameter settings are fixed as in the main case $Ra^*=50$, except for the order which is
set at $L=2$ to ensure good estimation quality of the transition properties across all OCs. 
The numerical values of the LOO test errors are reported in \cref{tab:errors_models_loo}. 
Higher errors are observed at the edge values of $Ra^*=30$ and $Ra^*=70$, which are extrapolation
cases. 

\begin{figure}[h!]
\centering
\includegraphics[width=\linewidth]{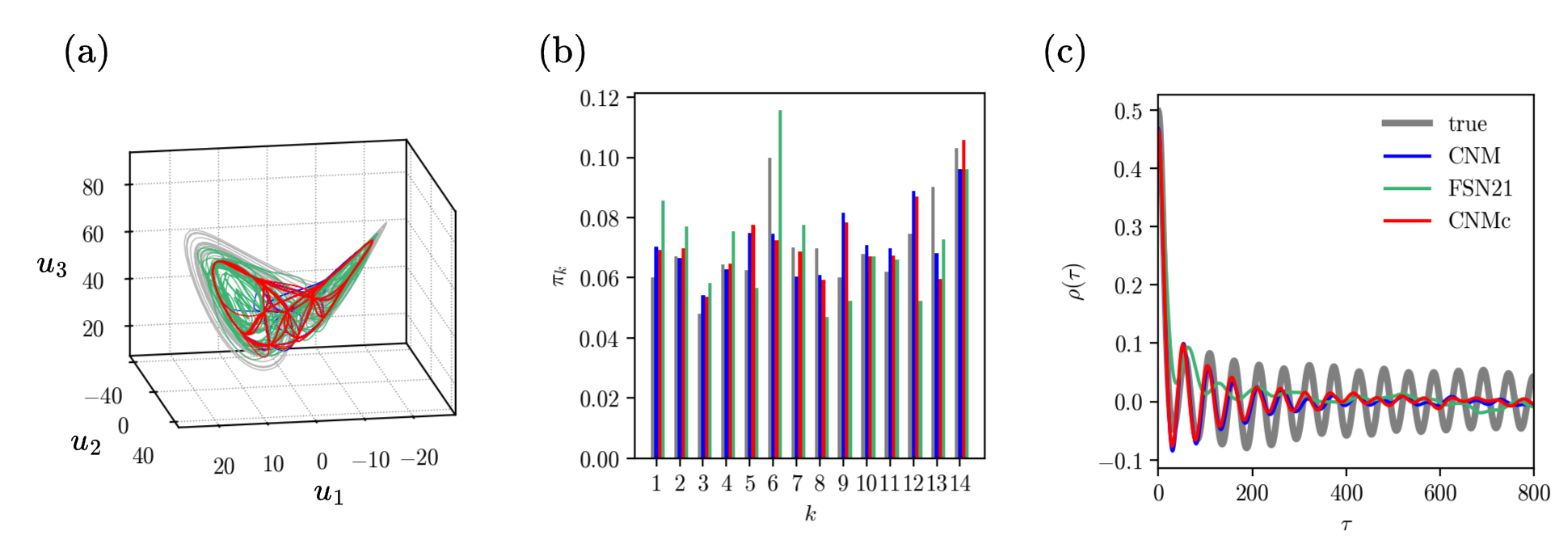}
\caption{
{\bf CNMc and FSN21 for Lorenz-63 at $Ra^*=50$.}
CNMc trajectories are compared to the reference CNM and the FSN21 model:
(a) state-space trajectories;
(b) stationary distributions;
(c) auto-correlation functions.
}
\label{fig:lorenz_results_fsn21}
\end{figure}

For completeness, CNMc is also benchmarked against the existing multi-OC cluster-based model
``FSN21'' \cite{Fernex2021a}.
Both models are trained with identical settings $K=14$ and $L=1$ (FSN21 does not support a
higher number of delays), and a comparison of the generated statistics is shown in
\cref{fig:lorenz_results_fsn21}.
CNMc outperforms FSN21, generating dynamics that are statistically nearly indistinguishable
from the reference CNM.
This conclusion is supported by the numerical values of the statistical discrepancies reported
in \cref{tab:errors_models_fsn21}.

\subsection{Controlled boundary-layer flow}\label{sec:controlled_tbl}

As a second case study, we consider a turbulent boundary-layer (BL) flow subjected to sinusoidal
wall corrugations that travel in the spanwise direction according to the time-dependent wall height
\begin{equation}\label{eq:waves}
y_\mathrm{wall}(z,t) = A\cos\!\left(\frac{2\pi}{\lambda} z - \frac{2\pi}{\Theta}t\right).
\end{equation}
The corrugation wavelength $\lambda$, wave period $\Theta$, and amplitude $A$ are the control
parameters that, taken together, define the operating condition.
All parameter values given below are expressed in viscous units.
This type of forcing generates a slow, two-dimensional, periodic unsteady flow component that
can reduce skin-friction drag for appropriate values of $\lambda$, $\Theta$, and $A$.
A comprehensive account of the flow physics and numerical simulation methodology is given in
\cite{Albers2020}.
As this system is governed by multiple parameters, inexpensive modelling of unseen OCs is
particularly valuable for reducing the number of expensive scale-resolved simulations needed to
characterise the flow.

Data are available at five non-overlapping combinations of $(\Theta,A)$, with a sixth
withheld as the test condition; the wavelength doesn't vary and is $\lambda=3000$ in all cases.
Likewise, the Mach number $Ma=0.1$, the momentum-thickness Reynolds number $Re=1000$ and the domain
size remain fixed.
The full-state data at each OC consist of $300$ snapshots of the three-dimensional flow field,
sufficient to span four of the slowest forcing periods $\Theta=141$.
Each snapshot comprises five $400 \times 100 \times 750$ scalar arrays, one per flow variable.
The data are prepared as follows to reduce them to a tractable size.


First, the slow unsteady flow component is extracted by averaging the full flow fields along the
streamwise direction and in phase with the wall forcing.
The field variables are then cast in a dimensionally consistent form through the energy-preserving
transformation 
\begin{equation}
(u,v,w,p,\rho) \mapsto (u\sqrt{\rho},\, v\sqrt{\rho},\, w\sqrt{\rho},\, \sqrt{\rho E}) = \mathbf{q},
\end{equation}
where $E$ is the fluid's specific internal energy and $\rho \approx 1$ owing to the negligible
compressibility of the flow \parencite{Rowley2004}.
The transformed snapshots $\mathbf{q}(t_i)$ from all OCs are then concatenated and subjected
to proper orthogonal decomposition to obtain a common empirical basis.
The normal form of the full-rank dynamics can be written in terms of the empirical mode coefficients
$\mathbf{a}$ as 
\begin{equation}\label{eq:dyn_sys_waves}
\frac{\rm d \mathbf{a}}{\rm d t} = \mathbf{F}(\mathbf{a},\, \mathbf{b}_{(\Theta,A)}).
\end{equation}
Finally, the snapshots from each OC are projected onto the empirical modes, and only the first
25 modes are retained, accounting for more than $98\%$ of the total variance across all OCs.
The resulting trajectories constitute the prepared dataset used to train the CNM models.

\begin{figure}[h!]
  \centering
  \includegraphics[width=\linewidth]{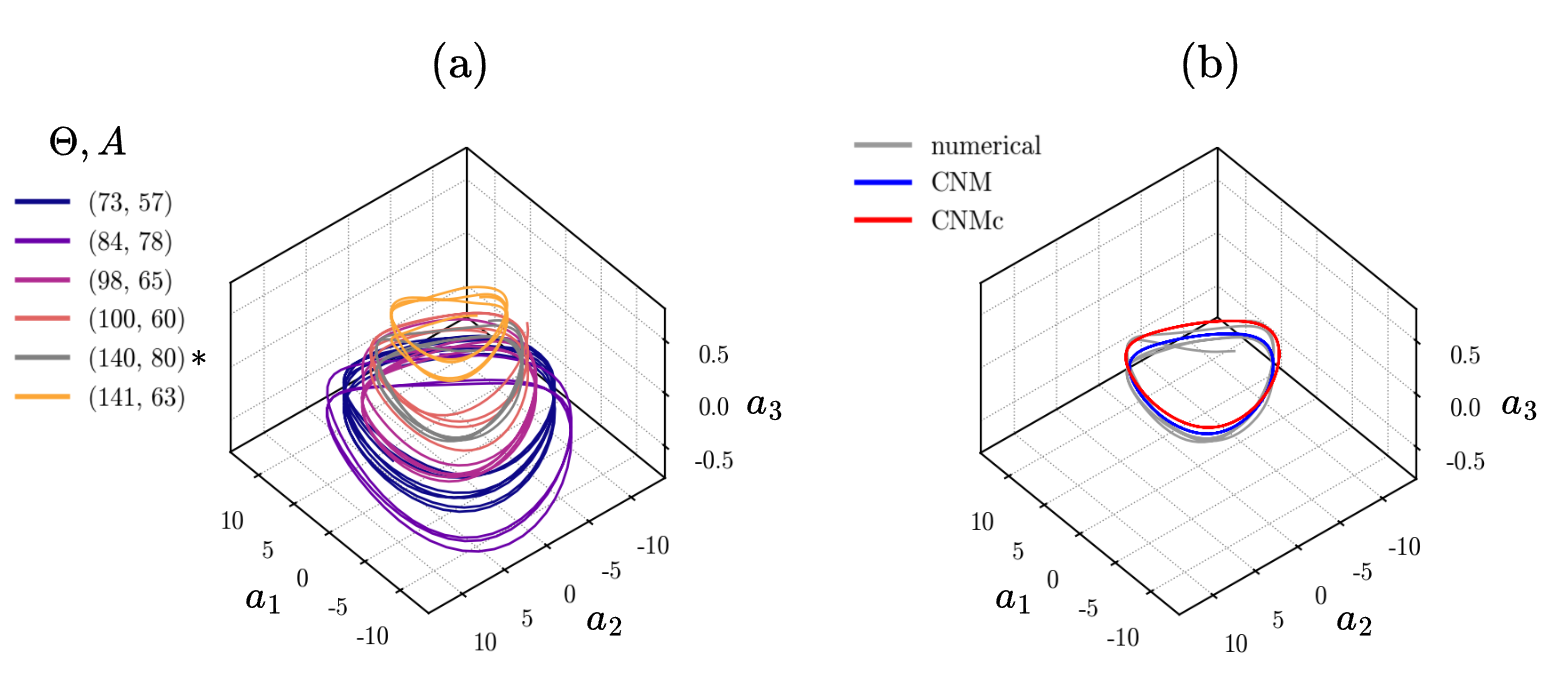}
  \caption{
  {\bf State-space trajectories of the boundary-layer flow.}
  (a) Numerically integrated trajectories; the test condition is indicated by ($\ast$).
  (b) Trajectories generated by the reduced-order models at the test condition.
  }
  \label{fig:bl_data}
\end{figure}

The trajectories truncated to the first three POD coordinates, still retaining more than
$97\%$ of the total variance, are displayed in \cref{fig:bl_data}a.
The trajectories appear as nearly regular orbits arranged on an approximately conical manifold
whose symmetry axis is aligned with the third coordinate $a_3$.
We note how the correlation between the position and radius of the orbits and the parameters
$\Theta,A$ appears to be non-trival. 

\begin{figure}[h!]
\centering
\includegraphics[width=\linewidth]{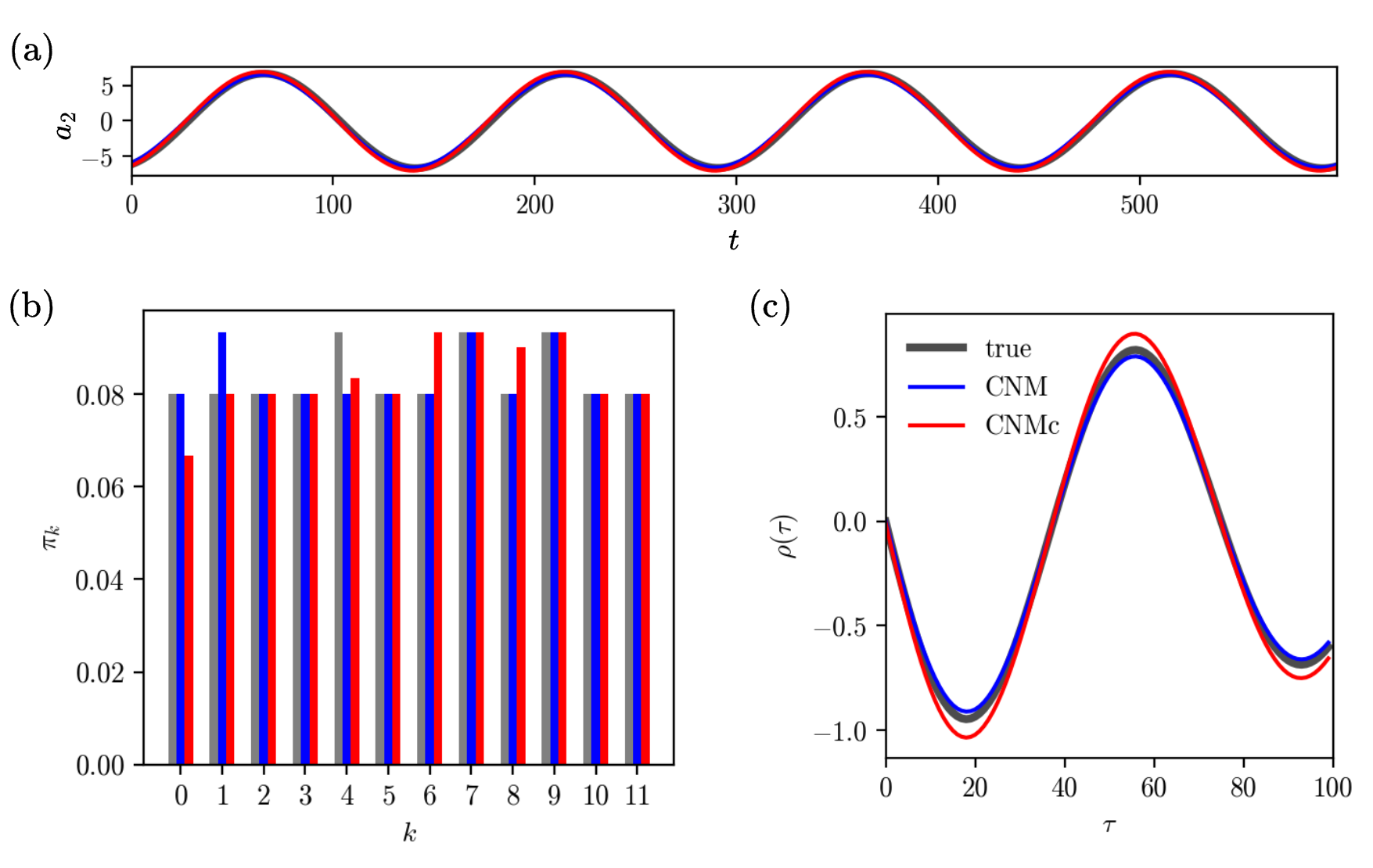}
\caption{
{\bf BL dynamics at the unseen condition $(\Theta, A)$.}
CNMc trajectories are compared to the reference CNM and the numerically integrated system:
(a) time-series of the $a_2$ component;
(b) stationary distributions;
(c) auto-correlation functions.
}
\label{fig:tbl_results}
\end{figure}

The CNMc algorithm is configured with $K=12$ cells and $L=1$ delay.
Linear regression models are used for both the transition properties and the parameters of the
transformation $\widehat{\phi}$, with hyperparameter selection carried out as in the Lorenz case.
The CNMc trajectory predictions shown in \cref{fig:bl_data}b and \cref{fig:tbl_results}
reproduce the period and amplitude of the reference CNM trained on the numerical data and also
closely mirror the numerically integrated trajectories.
The stationary histogram shows low discrepancy with the CNM reference, and the strong phase
coherence of the autocorrelation in \cref{fig:tbl_results} indicates that the transition times
were accurately inferred.
The autocorrelation also reveals a good match in amplitude between CNMc and the
CNM reference.

\begin{figure}[h!]
  \centering
  \includegraphics[width=\linewidth]{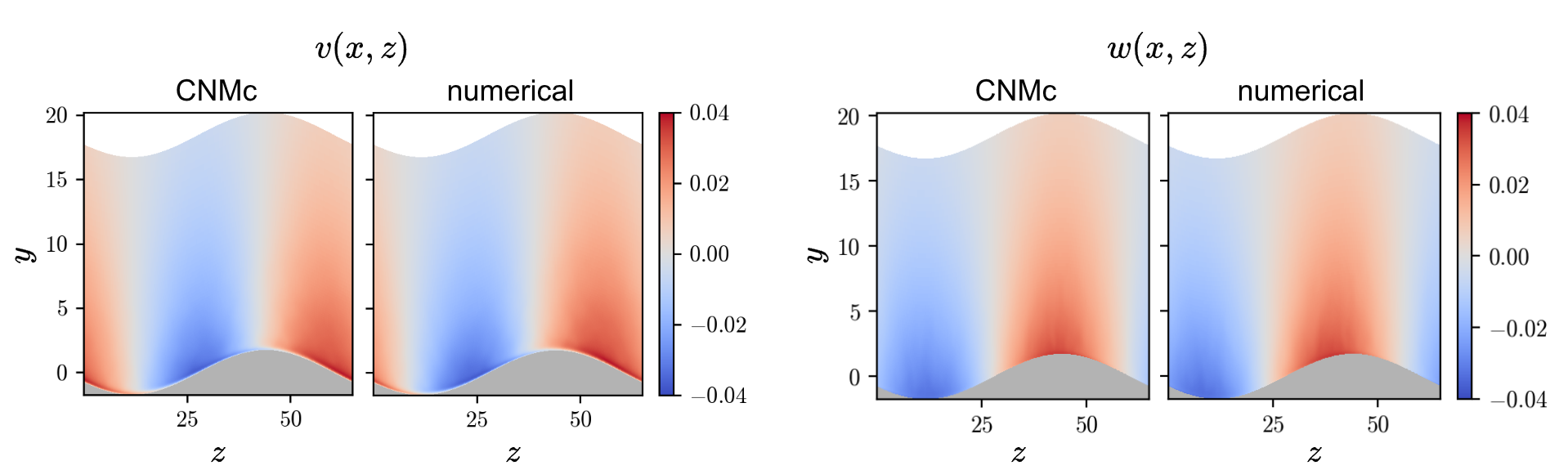}
  \caption{
  {\bf CNMc prediction of the unsteady flow component.}
  Snapshots of the unsteady velocity field at the test OC on a cross-flow section.
  }
  \label{fig:velocity_fields}
\end{figure}

The periodic nature of the unsteady flow component permits an instantaneous, in addition to a
statistical, assessment of the CNMc predictions.
\cref{fig:velocity_fields} shows one snapshot of the flow field predicted by CNMc at the unseen
OC, reconstructed in the natural variables $(u,v,w)$.
The CNMc trajectory is initialised from the same initial condition as the reference data, and the comparison is made after an elapsed time of $\Theta/4$.
The CNMc results and the numerical reference (also truncated to the first 25 modes) are in good agreement: only a slight phase lag is apparent, and the instantaneous velocity fields match
well visually.

\begin{figure}[h!]
\centering
\includegraphics[width=\linewidth]{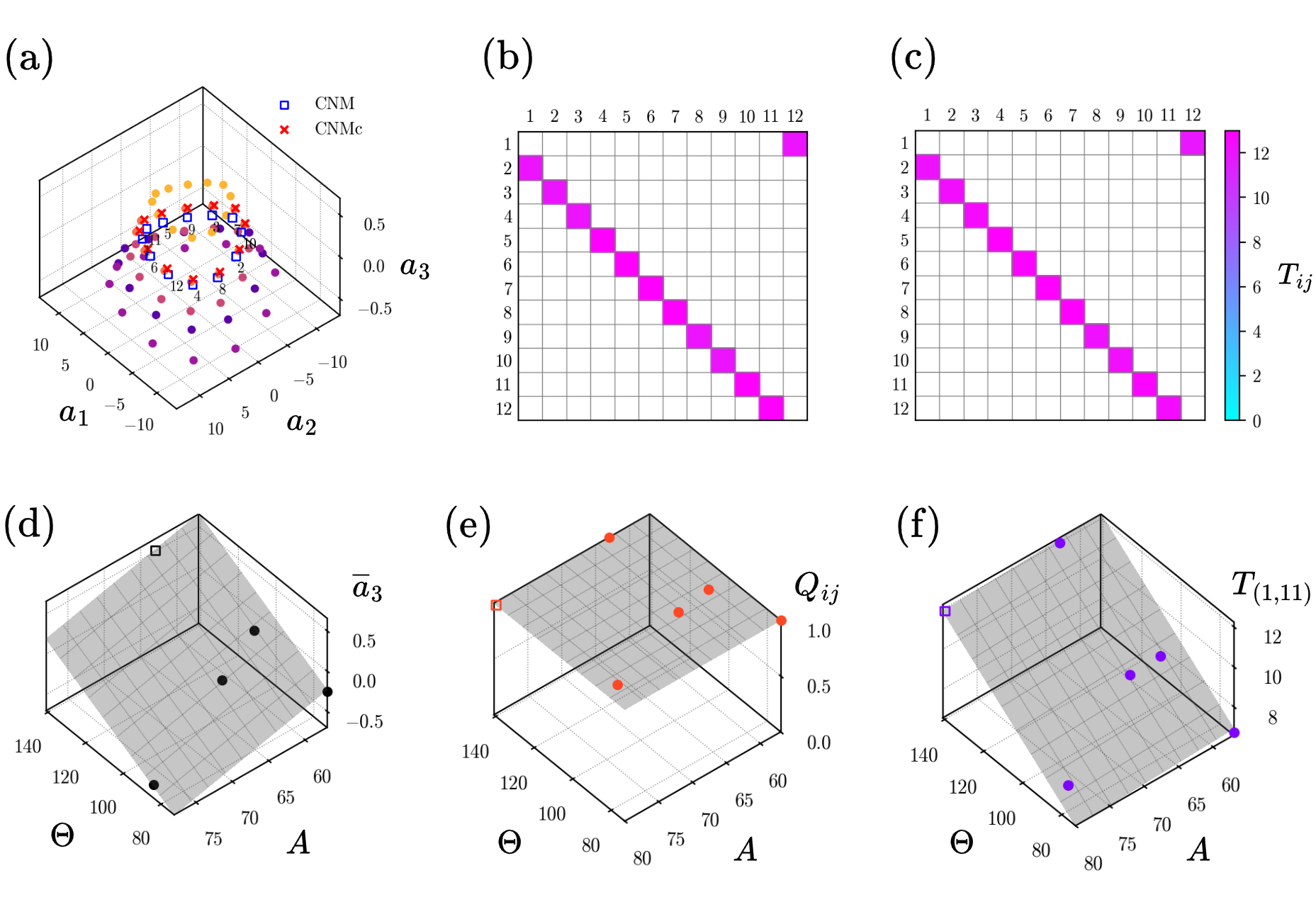}
\caption{
{\bf Elements of the CNMc model, BL case.}
(a) Predicted locations of the discrete states.
(b--c) Transition properties as predicted by (b) the CNM reference and (c) CNMc.
Regression models for
(d) the third component of the translation vector of $\phiocstar$,
(e) one element of $\Q{}$, and
(f) one element of $\T{}$.
The true value at the test OC is denoted by a square.
}
\label{fig:tbl_cnmc}
\end{figure}

Inspecting the model elements in \cref{fig:tbl_cnmc}, the predicted centroid positions in
\cref{fig:tbl_cnmc}a are generally accurate, though small positional errors plausibly
contribute to the amplitude discrepancy visible in the autocorrelations.
The transition matrices in \cref{fig:tbl_cnmc}b--c exhibit the same diagonal structure as the
CNM reference, correctly reflecting the periodic character of the numerical data.
The CNMc transition times also agree closely with the CNM reference, consistent with the
statistical results of \cref{fig:tbl_results}.
\cref{fig:tbl_cnmc}d--f depicts the bivariate regression models for one transformation
parameter and two elements of the transition matrices.
The linear regression models appear well suited to this system and generalise accurately to
the unseen OC.

\section{Summary and discussion}\label{sec:conclusions}
In this article, we have presented a general approach to generating out-of-sample reduced-order
global dynamics and developed a concrete algorithm that builds on the established technique of
cluster-based modelling.
The proposed CNMc algorithm operates in two stages.
First, it learns a common discrete state space across all observed regimes and estimates the
reduced-order dynamics on those common states via independent CNM models.
Then, it fits supervised regression models to the CROM parameters and to the state-space
transformation, enabling inference at operating conditions beyond the training range.
Trajectory generation at an unobserved regime then proceeds by predicting these parameters and
sampling from the resulting transition model, in the same fashion as standard CROMs.

The performance of CNMc was evaluated on two test systems, using dynamical trajectories at
sparsely sampled regimes for training and withholding one condition as the test case.
The statistics reconstructed by CNMc at the unseen regime are on par with those of a reference
CNM trained directly on the test data, demonstrating effective out-of-sample generalisation.
Under identical conditions, CNMc also displays greater robustness than the main competing
CROM-based method.
The controlled boundary-layer flow case successfully demonstrates the algorithm on a
multi-parameter system, a setting where avoiding costly scale-resolved simulations at every new
operating condition is of high practical value.

The experiments also reveal certain limitations of the current implementation of CNMc.
Most notably, its performance degrades beyond a certain number of time delays, which we
attribute to the growing estimation error on the CNM transition parameters as the number of
possible histories increases exponentially with the delay order.
CNMc also inherits the limitations of the base CNM method, most importantly its reliance on
high-order delays to capture non-Markovian dynamics and its susceptibility to data scarcity
in high-dimensional transition spaces.

Several avenues exist to address these shortcomings.
The dynamics-enhanced CNM variant \cite{Hou2023}, in particular, reduces the dependence of CNM
on high-order delays and could potentially also mitigate the estimation error that currently
limits CNMc at large $L$.
Another practically relevant scenario is one in which the available observations are corrupted
by noise.
We see no fundamental obstacle to applying the CNMc framework on top of a noise-robust CROM
as the base generative model in place of basic CNM.
Systems exhibiting more complex state-space evolution than those considered here may require
more expressive regression models than the simple ones employed above.
The choice of transformation class, in our case the Procrustes family, constitutes an
inductive bias that can in principle be generalised or tailored to specific systems whenever
information about their symmetries is available.
The fields of geometric machine learning and computational optimal transport offer tools that
may be well suited to the supervised modelling of graph-structured data and parametric
families of distributions that arise naturally in this context.
We are actively pursuing several of these directions.

{
\section*{Data availability}
A Python implementation of the algorithm is available at
\href{https://github.com/polivucci/cnmc}{https://github.com/polivucci/cnmc}.  
}

\begin{acknowledgments}


We would like to thank Xiao Shao and Prof. Wolfgang Schröder of RWTH Aachen for supplying the
numerical data of the turbulent boundary layer.
The research was funded by the Deutsche Forschungsgemeinschaft (DFG) in the framework of the
research project number 298994276.

\end{acknowledgments}

\bibliography{crom.bib}

@Article{Fernex2021,
  author   = {Daniel Fernex and Bernd R. Noack and Richard Semaan},
  journal  = {Sci. Adv.},
  title    = {Cluster-based network modeling{\textemdash}From snapshots to complex dynamical systems},
  year     = {2021},
  number   = {25},
  pages    = {eabf5006},
  volume   = {7},
  doi      = {10.1126/sciadv.abf5006},
  fjournal = {Science Advances},
}

@InProceedings{Fernex2021a,
  author    = {Daniel Fernex and Richard Semaan and Bernd Noack},
  booktitle = {{AIAA} Scitech 2021 Forum},
  title     = {Generalized Cluster-Based Network Model for an Actuated Turbulent Boundary Layer},
  year      = {2021},
  doi       = {10.2514/6.2021-1333},
}

@Article{Burkardt2006,
  author  = {John Burkardt and Max Gunzburger and Hyung-Chun Lee},
  journal = {{SIAM} Journal on Scientific Computing},
  title   = {Centroidal Voronoi Tessellation-Based Reduced-Order Modeling of Complex Systems},
  year    = {2006},
  number  = {2},
  pages   = {459--484},
  volume  = {28},
  doi     = {10.1137/5106482750342221x},
}

@Article{Kaiser2014,
  author   = {Eurika Kaiser and Bernd R. Noack and Laurent Cordier and Andreas Spohn and Marc Segond and Markus Abel and Guillaume Daviller and Jan Östh and Sini{\v{s}}a Krajnovi{\'{c}} and Robert K. Niven},
  journal  = {J. Fluid Mech.},
  title    = {Cluster-based reduced-order modelling of a mixing layer},
  year     = {2014},
  pages    = {365--414},
  volume   = {754},
  doi      = {10.1017/jfm.2014.355},
  fjournal = {Journal of Fluid Mechanics},
}

@Book{Lasota1994,
  author = {Andrzej Lasota and Michael C. Mackey},
  title  = {Chaos, Fractals, and Noise},
  year   = {1994},
  doi    = {10.1007/978-1-4612-4286-4},
}

@Article{Iacobello2022,
  author   = {Giovanni Iacobello and Frieder Kaiser and David E. Rival},
  journal  = {Phys. Fluids},
  title    = {Load estimation in unsteady flows from sparse pressure measurements: Application of transition networks to experimental data},
  year     = {2022},
  number   = {2},
  pages    = {025105},
  volume   = {34},
  doi      = {10.1063/5.0076731},
  fjournal = {Physics of Fluids},
}

@Article{Nair2019,
  author   = {Aditya G. Nair and Chi-An Yeh and Eurika Kaiser and Bernd R. Noack and Steven L. Brunton and Kunihiko Taira},
  journal  = {J. Fluid Mech.},
  title    = {Cluster-based feedback control of turbulent post-stall separated flows},
  year     = {2019},
  pages    = {345--375},
  volume   = {875},
  doi      = {10.1017/jfm.2019.469},
  fjournal = {Journal of Fluid Mechanics},
}

@Article{Kaiser2017,
  author   = {Eurika Kaiser and Bernd R. Noack and Andreas Spohn and Louis N. Cattafesta and Marek Morzy{\'{n}}ski},
  journal  = {Theor. Comp. Fluid Dyn.},
  title    = {Cluster-based control of a separating flow over a smoothly contoured ramp},
  year     = {2017},
  number   = {5-6},
  pages    = {579--593},
  volume   = {31},
  doi      = {10.1007/s00162-016-0419-4},
  fjournal = {Theoretical and Computational Fluid Dynamics},
}

@Article{Hou2023,
  author      = {Hou, Chang and Deng, Nan and Noack, Bernd R.},
  journal     = {arXiv},
  title       = {Dynamics-augmented cluster-based network model},
  year        = {2023},
  copyright   = {Creative Commons Attribution 4.0 International},
  doi         = {10.48550/ARXIV.2310.10311},
  eprint      = {2310.10311},
  eprintclass = {physics.flu-dyn},
  eprinttype  = {arXiv},
}

@Book{Bollt2013,
  author    = {Bollt, Erik M. and Santitissadeekorn, Naratip},
  publisher = {Society for Industrial and Applied Mathematics},
  title     = {Applied and Computational Measurable Dynamics},
  year      = {2013-11},
  isbn      = {9781611972641},
  doi       = {10.1137/1.9781611972641},
}

@Article{Souza2023,
  author      = {Souza, Andre N.},
  journal     = {arXiv},
  title       = {Transforming Butterflies into Graphs: Statistics of Chaotic and Turbulent Systems},
  year        = {2023-04},
  copyright   = {Creative Commons Attribution 4.0 International},
  doi         = {10.48550/ARXIV.2304.03362},
  eprint      = {2304.03362},
  eprintclass = {physics.flu-dyn},
  eprinttype  = {arXiv},
  publisher   = {arXiv},
}

@Article{Froyland2013,
  author    = {Froyland, Gary and Junge, Oliver and Koltai, Péter},
  journal   = {SIAM J. Numer. Anal.},
  title     = {Estimating Long-Term Behavior of Flows without Trajectory Integration: The Infinitesimal Generator Approach},
  year      = {2013-01},
  issn      = {1095-7170},
  number    = {1},
  pages     = {223--247},
  volume    = {51},
  doi       = {10.1137/110819986},
  fjournal  = {SIAM Journal on Numerical Analysis},
  publisher = {Society for Industrial & Applied Mathematics (SIAM)},
}

@Book{Guckenheimer1983,
  author    = {Guckenheimer, John and Holmes, Philip},
  publisher = {Springer New York},
  title     = {Nonlinear Oscillations, Dynamical Systems, and Bifurcations of Vector Fields},
  year      = {1983},
  isbn      = {9781461211402},
  doi       = {10.1007/978-1-4612-1140-2},
  fjournal  = {Applied Mathematical Sciences},
  issn      = {2196-968X},
  journal   = {Appl. Math. Sci.},
}

@Article{Lorenz1963,
  author    = {Edward N. Lorenz},
  journal   = {J. Atmos. Sci.},
  title     = {Deterministic Nonperiodic Flow},
  year      = {1963},
  number    = {2},
  pages     = {130 - 141},
  volume    = {20},
  doi       = {10.1175/1520-0469(1963)020<0130:DNF>2.0.CO;2},
  fjournal  = {Journal of Atmospheric Sciences},
  location  = {Boston MA, USA},
  publisher = {American Meteorological Society},
}

@Article{Kaiser2024,
  author    = {Kaiser, Frieder and Iacobello, Giovanni and Rival, David E.},
  journal   = {Philos. Trans. R. Soc. A},
  title     = {Cluster-based Bayesian approach for noisy and sparse data: application to flow-state estimation},
  year      = {2024-06},
  issn      = {1471-2946},
  number    = {2292},
  volume    = {480},
  doi       = {10.1098/rspa.2023.0608},
  publisher = {The Royal Society},
}

@Article{DeJesus2023,
  author    = {De Jesús, Carlos E. Pérez and Graham, Michael D.},
  journal   = {Physical Review Fluids},
  title     = {Data-driven low-dimensional dynamic model of Kolmogorov flow},
  year      = {2023-04},
  issn      = {2469-990X},
  number    = {4},
  pages     = {044402},
  volume    = {8},
  doi       = {10.1103/physrevfluids.8.044402},
  publisher = {American Physical Society (APS)},
}

@Article{Rowley2004,
  author    = {Rowley, Clarence W. and Colonius, Tim and Murray, Richard M.},
  journal   = {Physica D: Nonlinear Phenomena},
  title     = {Model reduction for compressible flows using POD and Galerkin projection},
  year      = {2004-02},
  issn      = {0167-2789},
  number    = {1–2},
  pages     = {115--129},
  volume    = {189},
  doi       = {10.1016/j.physd.2003.03.001},
  publisher = {Elsevier BV},
}

@Book{Holmes2012,
  author    = {Holmes, Philip and Lumley, John L. and Berkooz, Gahl and Rowley, Clarence W.},
  publisher = {Cambridge University Press},
  title     = {Turbulence, Coherent Structures, Dynamical Systems and Symmetry},
  year      = {2012-02},
  isbn      = {9781107008250},
  doi       = {10.1017/cbo9780511919701},
}

@Article{Kabsch1976,
  author    = {Kabsch, W.},
  journal   = {Acta Crystallographica Section A},
  title     = {A solution for the best rotation to relate two sets of vectors},
  year      = {1976},
  issn      = {0567-7394},
  month     = sep,
  number    = {5},
  pages     = {922--923},
  volume    = {32},
  doi       = {10.1107/s0567739476001873},
  publisher = {International Union of Crystallography (IUCr)},
}

@Article{Albers2020,
  author  = {M. Albers and P. S. Meysonnat and D. Fernex and R. Semaan and B. R. Noack and W. Schröder},
  journal = {Flow Turbul. Combust.},
  title   = {Drag Reduction and Energy Saving by Spanwise Traveling Transversal Surface Waves for Flat Plate Flow},
  year    = {2020},
  number  = {1},
  pages   = {125--157},
  volume  = {105},
  doi     = {10.1007/s10494-020-00110-8},
}

@Article{Cenedese2022,
  author    = {Cenedese, Mattia and Axås, Joar and Bäuerlein, Bastian and Avila, Kerstin and Haller, George},
  journal   = {Nature Communications},
  title     = {Data-driven modeling and prediction of non-linearizable dynamics via spectral submanifolds},
  year      = {2022},
  issn      = {2041-1723},
  month     = feb,
  number    = {1},
  volume    = {13},
  doi       = {10.1038/s41467-022-28518-y},
  publisher = {Springer Science and Business Media LLC},
}

@Article{Klus2018,
  author    = {Klus, Stefan and Nüske, Feliks and Koltai, Péter and Wu, Hao and Kevrekidis, Ioannis and Schütte, Christof and Noé, Frank},
  journal   = {Journal of Nonlinear Science},
  title     = {Data-Driven Model Reduction and Transfer Operator Approximation},
  year      = {2018},
  issn      = {1432-1467},
  month     = jan,
  number    = {3},
  pages     = {985--1010},
  volume    = {28},
  doi       = {10.1007/s00332-017-9437-7},
  publisher = {Springer Science and Business Media LLC},
}

\appendix
\section{Error criteria}
\label{app:statistical_criteria}

The quality of the predicted autocorrelation function $\widehat{\rho}(\tau)$ against a reference
$\rho(\tau)$ is assessed through the $L^1$ distance (or mean absolute error )
\begin{equation}\label{eq:mse_rho}
    \MAE(\rho, \widehat{\rho}) = \frac{1}{\| \rho(\tau) \|_1} \| \rho(\tau)-\widehat{\rho}(\tau) \|_1.
\end{equation}
The $L^1$  norm is calculated by integrating the argument between the positive and negative maximum
time lags. 
The autocorrelation function is calculated as
\begin{equation}\label{eq:autocorrelation}
    \rho(\tau) = \frac{1}{\mathrm{Var}(\mathbf{u}_{\rm data})} E_t(\mathbf{u}(t)\cdot\mathbf{u}(t-\tau)).
\end{equation}
where the normalisation factor is always the variance of the numerical data so as to allow a direct
comparison of their graphs. 

Analogously to the correlation MAE, the discrepancy between power spectral densities (PSDs) is defined by
\begin{equation}\label{eq:mse_psd}
    \MAE(S, \widehat{S}) = \frac{1}{\| S(f)\|_1} \| S(f)-\widehat{S}(f) \|_1
\end{equation}
where $f$ is the frequency and $L^1$ integral is calculated over the entire frequency range.

The model-generated discrete stationary distribution $\widehat{\boldsymbol{\pi}}$ is compared to the
reference $\boldsymbol{\pi}$, both being defined on the same state-space partition of $K$ cells.
The quality of $\widehat{\boldsymbol{\pi}}$ is assessed through the total variation (TV) distance
\begin{equation}\label{eq:tv_pi}
\TV(\boldsymbol{\pi}, \widehat{\boldsymbol{\pi}}) = \frac{1}{2}\sum^K_{k=1} |\pi_k - \widehat{\pi}_k|.
\end{equation}

\begin{table}[h!]
\caption{
Test errors of CNM and CNMc for Lorenz-63 at $Ra^*=50$ ($K=14$, $L=10$).
\label{tab:errors_models}
}
\begin{ruledtabular}
\begin{tabular}{lllll}
model  & reference & $\TV(\boldsymbol{\pi}, \widehat{\boldsymbol{\pi}})$ & $\MAE(\rho,
\widehat{\rho})$ & $\MAE(S, \widehat{S})$   \\
\midrule
CNM    & data      &  0.07  &  0.42  &  0.30  \\
CNMc   & data      &  0.07  &  0.57  &  0.32  \\
CNMc   & CNM       &  0.02  &  0.20  &  0.16  \\
\end{tabular}
\end{ruledtabular}
\end{table}

\begin{table}[h!]
\caption{
Test errors of CNM, CNMc and FSN21 for Lorenz-63 at $Ra^*=50$ ($K=14$, $L=1$). 
\label{tab:errors_models_fsn21}
}
\begin{ruledtabular}
\begin{tabular}{lllll}
model  & reference & $\TV(\boldsymbol{\pi}, \widehat{\boldsymbol{\pi}})$ & $\MAE(\rho,
\widehat{\rho})$ & $\MAE(S, \widehat{S})$   \\
\midrule
CNM    & data      &  0.07  &  0.72  &  0.35 \\
CNMc   & data      &  0.07  &  0.78  &  0.35 \\
CNMc   & CNM       &  0.01  &  0.57  &  0.19 \\
FSN21  & data      &  0.09  &  0.94  &  0.46 \\
FSN21  & CNM       &  0.11  &  0.97  &  0.44 \\
\end{tabular}
\end{ruledtabular}
\end{table}

\begin{table}[h!]
\caption{
Leave-one-out test errors of CNMc relative to CNM, Lorenz-63 ($K, L = 14, 2$).
\label{tab:errors_models_loo}
}
\begin{ruledtabular}
\begin{tabular}{llll}
Leave-out $Ra^*$ & $\TV(\boldsymbol{\pi}, \widehat{\boldsymbol{\pi}})$ & $\MAE(\rho, \widehat{\rho})$ & $\MAE(S, \widehat{S})$   \\ 
\midrule
30 & 0.07 & 0.79 & 0.43 \\
40 & 0.04 & 0.37 & 0.16 \\
50 & 0.05 & 0.30 & 0.12 \\
60 & 0.05 & 0.37 & 0.16 \\
70 & 0.08 & 1.01 & 0.32 \\
\end{tabular}
\end{ruledtabular}
\end{table}



\section{Influence of hyperparameters $K$ and $L$}
\label{app:lorenz_hyperparameter}

\begin{figure*}[ht]
\centering
\includegraphics[width=0.9\linewidth]{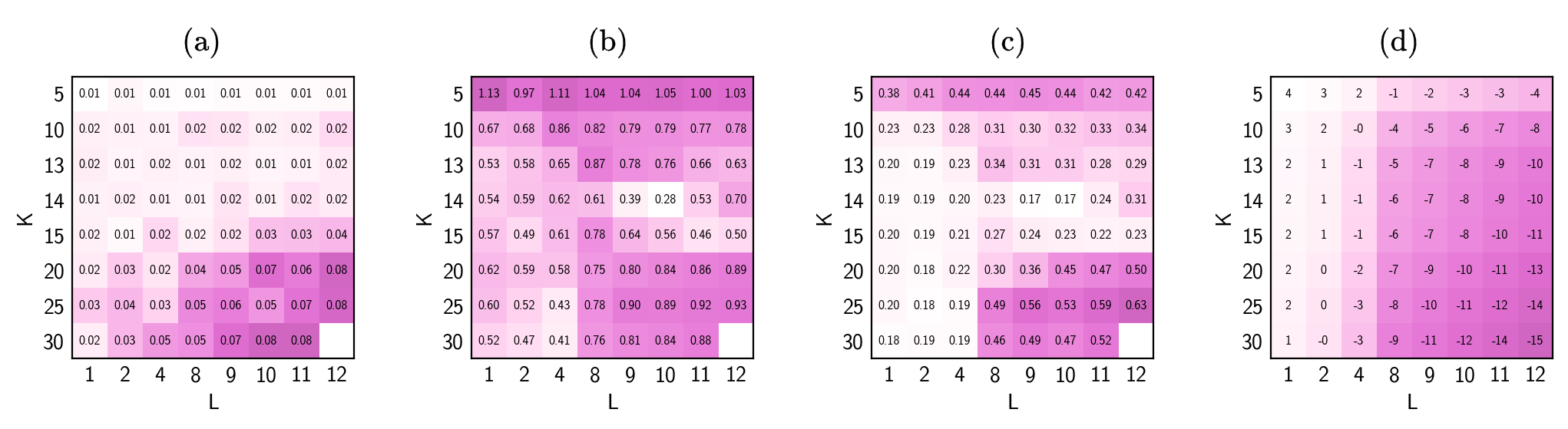}
\caption{
{\bf Influence of hyperparameters $K$, $L$ on the test error, Lorenz data.}
Statistical discrepancies between CNMc and CNM at the test OC for 63 combinations of the
hyperparameters $K,\,L$: 
(a) histogram TV distance;
(b) autocorrelation MAE; 
(c) spectral MAE;
(d) estimated amount of data available per entry of $\mathbf{Q}$, in $\log_{10}$ units. 
The values at $(30,12)$ are missing owing to numerical limitations.
}
\label{fig:lorenz_grid_cnmc_vs_cnm}
\end{figure*}
In \cref{sec:lorenz_system} we remark how the statistical accuracy of CNMc drops at high values of
$L$, contrary to what holds true for CNM. 
\cref{fig:lorenz_grid_cnmc_vs_cnm}a-c look more closely at the discrepancy between CNMc and CNM.
The test errors of both the stationary distribution and the time-series degrades at high values of
$K$ and $L$. 
In the text, we conjecture that this phenomenon stems from the increasingly large estimation error
associated with calculating the probabilities of ever rarer transitions from finite-length data.  
The supervised model would then fit this very small sample and predict spurious transitions.
A crude estimate of the average amount of data available for each entry of $\Qhat$ is given
by $T_K / K^{L+1} \propto 1/K^{L+2}$ with $T_K=E(T_k)/T$ being the average holding time in a cell
expressed as a fraction of the total data time (here $T=500,000$ time steps). 
The pattern visualised in \cref{fig:lorenz_grid_cnmc_vs_cnm}d, while not conclusive evidence, lends
qualitative support to this hypothesis.  

In addition the performance drop at high $K$ and $L$, \cref{fig:lorenz_grid_cnmc_vs_cnm}b-c shows
that the auto-regressive statistics are also uniformly inaccurate at very low cell counts ($K=5$), a 
fact that is not observed for the histogram. 
A similar low-$K$ inaccuracy affects the CNM statistics with respect to the data (not shown).
At present we do not have an hypothesis on why this is the case.

These two limitations cause the overall more accurate CNMc models to be found at moderate
values of $K$ and $L$; the best combination ($14, \, 10$) belongs to this range.

\end{document}